# Staying vigilant in the Age of AI: From content generation to content authentication

**Yufan Li**
Computational Media and Arts
Hong Kong University of Science and Technology (Guangzhou)
Nansha 511458, China
yli538@connect.hkust-gz.edu.cn

**Zhan Wang**
Computational Media and Arts
Hong Kong University of Science and Technology (Guangzhou)
Nansha 511458, China
zwang834@connect.hkust-gz.edu.cn

**Theo Papatheodorou**\*
Computational Media and Arts
Hong Kong University of Science and Technology (Guangzhou)
Nansha 511458, China
theodoros@hkust-gz.edu.cn

**Abstract**

This paper presents the Yangtze Sea project, an initiative in the battle against Generative AI (GAI)-generated fake content. Addressing a pressing issue in the digital age, we investigate public reactions to AI-created fabrications through a structured experiment on a simulated academic conference platform. Our findings indicate a profound public challenge in discerning such content, highlighted by GAI's capacity for realistic fabrications. To counter this, we introduce an innovative approach employing large language models like ChatGPT for truthfulness assessment. We detail a specific workflow for scrutinizing the authenticity of everyday digital content, aimed at boosting public awareness and capability in identifying fake materials. We apply this workflow to an agent bot on Telegram to help users identify the authenticity of text content through conversations. Our project encapsulates a two-pronged strategy: generating fake content to understand its dynamics and developing assessment techniques to mitigate its impact. As part of that effort we propose the creation of speculative fact-checking wearables in the shape of reading glasses and a clip-on. As a computational media art initiative, this project under-scores the delicate interplay between technological progress, ethical considerations, and societal consciousness.

## Introduction

The proliferation of generative content is increasingly evident in various aspects of our world, with a growing number of researchers delving into algorithmic studies to enhance the quality of generated text, images, audio, and more. Amid this surge in research activity, it is prudent to pause and consider the necessity and implications of studying and employing generative AI technology.

## Generative AI and its Positive Side

Upon examining the historical trajectory of generative AI, we observe that initial studies of generative models were primarily focused on comprehending and modeling the structure and distribution of data [1] [2]. The primary objective was to generate novel data samples that mirrored the training data, thereby enabling us to bridge data gaps and undertake data augmentation.

For instance, in the case of class-imbalanced datasets, where there is a disproportionate volume of data under a specific classification, the generative model can be employed to supplement the deficit of data. This results in a more balanced and robust dataset for subsequent training [3].

The utility of this data generation technique is especially pronounced in medical and pharmaceutical research. Given that these fields often involve data of a highly sensitive and private nature, these algorithms can generate simulated data based on the original dataset for subsequent research, thereby maintaining confidentiality [4] [5].

At the same time, the potential of generative AI is being explored in the domain of creative content production [6] [7]. There is a palpable enthusiasm surrounding the use of generative AI, which enables individuals to effortlessly create seemingly high-quality text, images, and songs. Some argue that academic institutions or corporations studying these algorithms are essentially democratizing innovation by lowering the barriers to content creation [8].

## Ethical Concerns of generative models

Technical literature abounds with commendations for the myriad positive facets of generative AI, but how many studies truly consider its potential pitfalls? Regrettably, the examination of critical ethical elements is often overlooked. A comprehensive literature review of 884 papers in the domain of generative audio models revealed that a mere 10% of these studies contemplate the potential negative impacts or identify types of ethical implications [9].

Recently, a research group demonstrated that in controlled laboratory conditions only 68% of generative content was correctly identified as such by human domain experts

[10]. It is cause for concern that nearly one-third of fake articles generated are not detected by top reviewers.

An early attempt to generate a philosophical essay in a computer-simulated postmodernist style was submitted and accepted by Philosophy and Literature, a prominent American literature journal [11]. This incident, known as the Sokal Affair, ignited a debate about philosophical and social science essay writing of the period.

Moreover, there is an art project that have proposed and technically discussed pipelines and techniques for the wholesale generation of fake news [12]. It explores how machine learning methods can be used to generate fake news and present them in the guise of an online news blog. Recently, a notable art news story emerged: an AI-generated image won a prestigious international photography competition[1]. Such fake content directly challenges us to reconsider the implications of generative techniques.

Our work is an extension of these prior experiments. Our primary focus lies in understanding the real-world impact of the generated content. We aim to unravel the workings behind these phenomena and explore how we can guard against dangerous generative content with generative models. We propose a novel approach for truthfulness assessment and introduce a workflow for evaluating the truthfulness of everyday digital content.

# Art Practice: Human Reactions to Synthetic Fake Content

## Setup

We originally created fake content using ChatGPT and Midjourney, and hosted it on a website designed to mimic an academic conference. This site was linked to popular social media platforms.

**Generation.** Starting with a real archaeological discovery [13], we generated five fake papers complete with titles, authors, and abstracts. These papers featured fictional discoveries like colossal dragons and connections between ancient civilizations. We also set up a fake academic conference website with details like conference name, schedule, open call for submissions, program session and committee members. All the fake content can be found in our website[2].

**Assembly and interaction.** Our website, shown in Figure 1 (left), resembled a standard academic conference called Chinese Archaeology and Cultural Research (CACR2023). The site included hidden "Easter Eggs" that revealed the generative process behind the content. Clicking on any part of the site displays screenshots of our interactions with ChatGPT or Midjourney. An apology letter explaining our project is carefully hidden in the "contact us" section. While the site appears typical at first glance, deeper exploration reveals its generative nature, as shown in Figure 1 (right).

**Distribution Process.** To test public reaction, we shared our fake papers and website on platforms frequented by our target audience. The distribution was twofold. For experts, we directly emailed ten archaeology scholars specializing in ancient China, sharing the paper's abstract and the conference website, inviting them to review.

For the general public, we posted the content on Wikipedia and Twitter, and raised discussions on Quora and Zhihu (a Chinese Q&A platform where questions are posted and answered). This approach aimed to elicit a broad range of responses from a diverse audience, who received this information indirectly through social media.

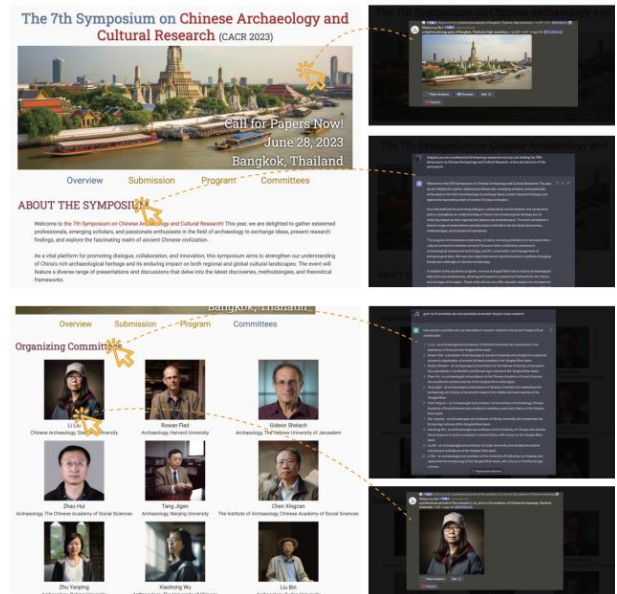

Figure 1. The fake academic conference. Clicking on any part of the site displays screenshots of our interactions with ChatGPT or Midjourney, showing how we generated the content.

## Responds

We documented responses from experts and the general public in Table 1, totaling 36, with 15 correctly identifying our simulated content. Overall, 42% detected our deception. However, few explored the website's interactive elements revealing our methods. All responses and links are available on our website.

**Expert Responses.** Of the 10 expert responses, 70% showed no interest in our manuscripts or the conference, among which 30% remarked on the unusual nature of the content. Intriguingly, two experts expressed significant interest and willingness for collaboration. This contrast in expert engagement highlights varied levels of skepticism and openness within the academic community.

**General Public Responses.** Most public responses came from Quora, where 41% recognized our deception. We posed questions about our fake discoveries, receiving varied reactions, from high praise to skepticism, as shown in Table 2. Interestingly, some supportive responses, particularly those elaborating on our pseudo-discoveries, seemed to be generated by tools like ChatGPT or other LLMs (see the answers in this link[3]). On Zhihu, we received two responses: one identified our ruse, and the other, possibly AI-generated (seen here[4]), provided detailed insights into our fabricated findings.

---

[1] https://reurl.cc/xLlmb5

[2] http://www.cacr-symp.com/

| | Number of responds | Spot the tricks | Positive responds | Negative responds |
|---|---|---|---|---|
| Experts - Direct information | | | | |
| Email | 10 | 3 | • Interested in reviewing the work, "There has been considerable work on Yangzi River valley cultural finds of Neolithic date."<br>• Willing to attend the conference<br>• "Delighted to hear about the new archaeological discoveries in the Yangtze River Basin." | • "Empty. Filled with big words, but nothing specific."<br>• "Very strange topic."<br>• Do not want to review the manuscripts. |
| General people - Second hand information | | | | |
| Quora | 22 | 9 | • Gave more evidences about the civilization, culture, or creatures developed/discovered along Yangtze River.<br>• "Provides important insights"<br>• "Highlights the complex interplay"<br>• "Interesting in this case"<br>• "This shed new light on the Marine life and biodiversity of the Late Triassic Period." | • "Based on current scientific knowledge, this is not possible."<br>• "I couldn't find any newly discovered Art in the Yangtze river basin."<br>• "You're under the influence of ayahuasca."<br>• "There cannot be a cultural connection"<br>• "There was no Triassic period." |
| Zhihu | 2 | 1 | • Gave more evidences about the culture and art developed/discovered along Yangtze River. | • Directly spotted the tricks |
| Twitter | 0 | 0 | / | / |
| Wikipedia | 2 | 2 | / | Deleted the post and banned the account. |

Table 1. Responds from experts and general people.

This phenomenon suggests that generative tools are not only used for creating fake content but also for responding to it, blurring the lines between human and AI-generated reactions. This stark contrast between the high praise to skepticism responses underscores the varying degrees of critical engagement by the audience.

Our Twitter posts saw minimal engagement, with no responses. Wikipedia quickly deleted our content and banned our account. All in all, the responds form the public and actions by platform reflect the differing levels of vigilance and moderation policies across platforms.

**Initial Findings and Inspirations**

Our analysis led to several key findings:
1. **Public Awareness and Detection Skills**: There's a noticeable gap in public awareness and overestimation of our ability to detect artificially generated content. Most people, including experts, struggle to identify its artificial nature.
2. **Second-hand Information Risks**: The use of deceptive websites for information dissemination highlights the vulnerability of individuals who don't verify information sources, leading to easy misinformation spread.
3. **Democratization of Deception**: The rise of GAI has lowered the barriers for spreading false information, marking a shift towards the democratization of deception, unlike the positive connotations associated with the democratization of innovation.
4. **Text vs. Image Deception**: Generated images are more easily identified as fake compared to text. Text generation is more deceptive than that of image generation.
5. **Self-Perpetuating Cycle of AI**: The widespread use of tools like ChatGPT suggests a cycle where algorithms generate content and responses, diminishing the human role to mere information transmitters.

The key issue we've identified is that Generative AI has greatly lowered the threshold for creating false information. While the quality of such generated content has improved, making it more convincing, there's a concerning lag in public awareness and the availability of tools for detecting fake content. This growing disparity poses a significant challenge.

To tackle this, in the second part of our performative experiment we created a prototype pipeline to fact-check statements using LLMs. Our goal is to counter the widespread ease of creating deceptive content. We speculate on the use of accessible solutions that help people recognize not just AI-generated, but all kinds of fake content in their daily lives.

---

[3] https://reurl.cc/M4RQmm

[4] https://www.zhihu.com/question/601576432

## Emphasizing Reasoning Over Detection

Currently, the field of generative AI is experiencing a surge of interest, leading to an adversarial research environment between generative and detection mechanisms.

On one hand, an increasing number of generative AI models are being developed. The objective of these algorithms is not merely to generate creative content, but to produce output of such quality that it could surpass the creative capabilities of human experts.

Conversely, numerous algorithms strive to accurately identify artificially generated content to prevent humans or systems from being deceived. These include algorithms for detecting generated images [14], Twitter posts[15], news articles [16], and even fingerprints [17], etc. The primary objective of these detection algorithms is to ensure data integrity.

The current situation indicates that generative AI's capabilities exceed those of detection algorithms, primarily based on Machine Learning, Deep Learning, and Natural Language Processing [18]. This is largely due to the fact that the outputs of generative AI have become more universal, applicable to a broader array of tasks, and increasingly leaning towards general intelligence.

## Methods

Consequently, rather than focusing on creating a new detection model for a specific type of content, we decided to approach this differently. Instead of detecting, we aimed to reason about the veracity of given claims. Essentially, we planned to have the general intelligence model like ChatGPT determine whether a given statement aligns with facts or logic, and subsequently assign a truth score. To test the feasibility of this idea, we experimented with GPT4 with web plugins, fine-tuned GPT, and Agent GPT. Agent GPT refers to a variant of the GPT model designed for interactive and autonomous tasks. This autonomous GPT model can perform more complex operations like browsing the web or using tools to gather and process information in real-time. Efforts to employ Large Language Models (LLMs) for fact-checking have demonstrated their potential in this area [19]. Additionally, empirical research on using LLMs for fact verification has highlighted both the risks and opportunities associated with this approach [20].

We aim for GPTs to not only provide a veracity score through reasoning but also offer reasons and identify suspicious parts of a given statement. To achieve this, we employ prompt engineering, specifically prompting with a few shots, to provide specific instructions to GPT.

**GPT4 with web access plugin.** We designed a four-part prompt consisting of a role set that establishes GPT's role as an expert professor capable of discerning lies, and an instruction section outlining the tasks and corresponding rules for GPT4, few-shot examples, and the input question (Figure 2). We utilized the web plugin feature of GPT, enabling it to access more current news and reason with this information, thus overcoming the limitations imposed by the model's "cut-off date.". The first task for GPT is to determine if a statement is objective or verifiable. Only if the statement is verifiable will GPT provide a specific veracity score, identify suspicious parts of the statement, and provide corresponding reasons. We provided four examples to aid GPT in understanding the formatting norms for potential cases.

**Agent GPT.** In addition to the mentioned prompt, we integrated the suggested prompt[5] for building Agent GPTs. By doing so, we enabled the original GPT4 to actively participate in the process of breaking down the task of determining true and false claims into smaller, manageable tasks. These tasks are executed step-by-step, utilizing different plugins like Bing search, to gather relevant information and navigate through the process. This approach allows the GPTs to systematically tackle the task and obtain a final response through a series of well-defined steps and the utilization of external resources.

Figure 2. The prompt for GPT4 with web plugins.

| Methods | Total statements | Correct answers | Wrong answers | Unable to Judge | Accuracy |
|---|---|---|---|---|---|
| GPT4 with Few-shot Prompts | 20 | 14 | 3 | 3 | 82% |
| Agent GPT | 20 | 16 | 3 | 1 | 84% |

Table 2. Veracity assessment accuracy for three methods.

---
[5]https://reurl.cc/Z9Xkmp

**Fine-tuned GPT** is an approach that is tailored to individual cases. In this approach, we fine-tuned the GPT model on specific datasets and tasks related to determining the veracity of information. By fine-tuning the model on these specific cases, we aimed to enhance its ability to accurately differentiate between true and false claims. This case-by-case fine-tuning approach allowed us to adapt the GPT model to the specific requirements of the task at hand, improving its performance and reliability in assessing the veracity of information.

| Original Statement | Ground Truth | Method | Veracity | False Part | Reason |
|---|---|---|---|---|---|
| MoveOn.org says "McCain opposes a woman's right to choose." | ✅ mostly-true | GPT4 | ✅ mostly-true | / | John McCain's public record and statements have consistently shown that he was antiabortion, supporting the overturning of Roe v. Wade and favoring a Constitutional amendment to ban abortion, with exceptions only in cases of rape, incest, or threat to the life of the mother. |
| | | Agent GPT | ✅ true | / | It is accurate to say that McCain was pro-life and publicly stated his opposition to the Roe. vs. Wade decision, advocating for its overturn. This position was also confirmed on his campaign website, where it was stated that he would nominate judges who shared this viewpoint. |
| Mitt Romney says that illegal immigrants get a $100,000 break on University of Texas tuition over four years. | ✅ mostly-true | GPT4 | ❌ mostly-false | $100,000 break | The claim that illegal immigrants receive a $100,000 break on tuition at the University of Texas is not supported by the available data. While there were provisions for in-state tuition for certain undocumented students, the specific figure of a $100,000 tuition break over four years is not substantiated by the sources. |
| | | Agent GPT | ❌ mostly-false | $100,000 break over four years | While Mitt Romney accurately cited that illegal immigrants receive a significant in-state tuition discount at the University of Texas, the exact amount of nearly $100,000 over four years is slightly exaggerated. The actual calculated difference based on the 2011-12 tuition charges would be $90,800, not $100,000. Furthermore, t Only 4% of the illegal immigrants benefiting from in-state tuition attended UT. The majority chose community colleges, receiving much smaller tuition reductions averaging $1,600 to $2,600 annually. |
| Facebook posts "The New York Times published an old stock photo of a young girl and claimed Israeli forces killed her during its recent war with Hamas. | ✅ mostly-true | Agent GPT | ❌ false | $100,000 break over four years | The claim that The New York Times published a stock photo of a young girl and falsely reported that Israeli forces killed her during a conflict with Hamas is not supported by evidence. The New York Times has a rigorous editorial process, and such a significant error would have been widely reported and corrected. |
| | | Agent GPT | ❌ false | The New York Times intentionally used an old photo claiming Israeli forces killed the girl recently | The New York Times did mistakenly use an old image of a girl who was not killed, but the error was due to human error and not an attempt to deceive. The Times issued a correction and replaced the photo with the correct one provided by the family of the deceased |

Table 3. Three examples of false part and reason from GPT4 with web plugin and Agent GPT.

## Evaluation and Results

We conducted experiments using the same dataset to evaluate the two methods and assessed the accuracy of their judgments. Since existing true-false information datasets primarily focus on news, we selected a true-false news dataset from Kaggle[6]. We randomly chose 20 news articles from this dataset for testing and obtained the following results.

Based on the overall results, the accuracy of the fine-tuned GPT is much lower than the other two methods, whereas both GPT4 and Agent GPT achieved similar performance with over 80% correct judgments (shown in Table 2). The errors identified by these methods were also relatively similar. There are three pieces of data that cannot be evaluated for GPT4, while only one cannot be judged by Agent GPT. The reasons provided for the inability to assess these data are reasonable. When determining the final accuracy rate, we only take into account the news that can be verified as true or false. However, when it comes to detailed inferences, Agent GPT provided more reasonable justifications for its judgments. It is important to note that the inference screening process of Agent GPT is more rigorous and time-consuming, resulting in longer running times.

Table 3 presents three representative examples. The first example represents the majority of news articles that were accurately assessed, as both GPT4 and Agent GPT provided correct answers along with sound justifications. However, the second and third examples are more unique in that one of the methods yield different answers compared to the Ground Truth in each example, yet still offer reasonable explanations.

In the second example, the original news investigation article states that the state government reduces tuition fees for local residents and allows immigrants with local accounts to benefit from these reductions. GPT4, however, believes that there is no direct policy supporting this claim, leading to a reasonable judgment of its inaccuracy. On the other hand, Agent GPT argues that while the original description is correct, it is not representative of the overall situation, providing a detailed explanation of the limited number of immigrants at the University of Texas.

Moving on to the third example, The New York Times inaccurately reported an event that did occur. GPT argues that the event was reported by The New York Times, as the judgmental statement is accurate. On the other hand, Agent GPT conducts a more in-depth analysis and deduces that there may be subsequent news clarifications from The New York Times regarding the misreporting, leading to the judgment that the news is false.

In summary, considering the aforementioned analysis, we believe that among the two methods proposed in this paper, GPT4 with the web plugin is more suitable for practical applications in veracity assessment. It strikes a balance between accuracy and efficiency, making it a more practical choice for real-world scenarios.

# Prospective Usage: Assessing Veracity in Everyday Content

## Workflow Design

The outcomes of our previous research confirm that large language models are more effective for reasoning through content than traditional fake news detection methods. With this in mind, we have developed a clear, step-by-step workflow to check the veracity of the digital content we see every day. You can find this process illustrated in the left part of Figure 3.

Every day, people interact with a mix of text, video, and audio information online. For web page texts, we can directly extract the content. When it comes to videos and audio, we can use subtitles or convert the audio into text. We would then feed this text to ChatGPT, sentence by sentence, to evaluate its veracity using the custom inference prompt presented earlier. ChatGPT would then provide a score (local veracity score) for each sentence's veracity, identify false elements, and explain the reasoning. We also compute a total veracity score for the entire text. This score evaluates the overall truthfulness of the content currently displayed, such as the text visible on a screen page or the portion of a speech heard in a video up to that point. It provides a comprehensive assessment of the content you are currently experiencing.

We then relay the sentence's veracity score, the identified false parts, and the Global Veracity score back to the original web page or video. We visualize these results with a bar chart and red dashes, as shown in the right part of Figure 3. Sentences with questionable parts are marked with red underlines, and the veracity score is displayed in the top left corner.

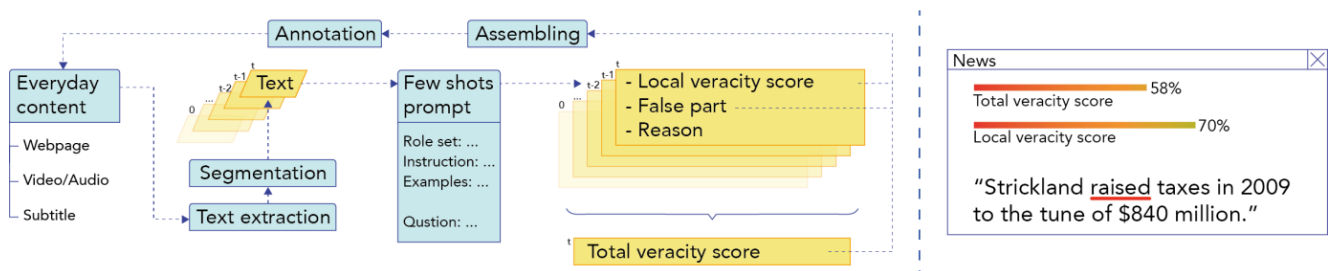

Figure 3. The workflow of assessing veracity in everyday content

---

[6] https://www.kaggle.com/datasets/rmisra/politifact-fact-check-dataset/

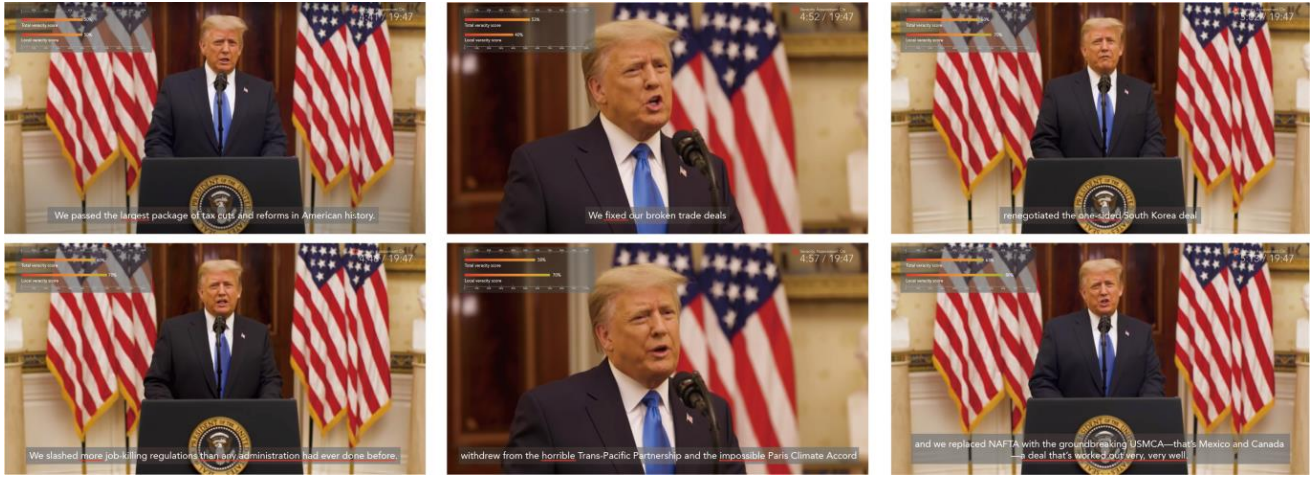
Figure 4. Veracity assessment workflow applying to the Farewell Address of President Donald J. Trump.

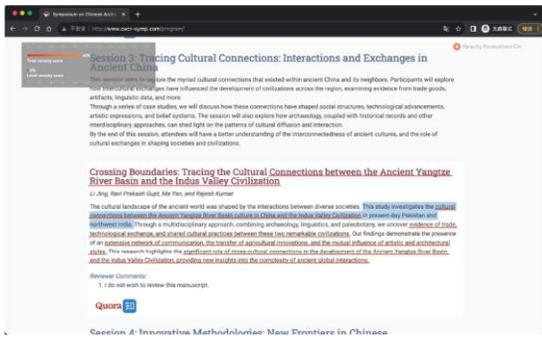
Figure 5. Veracity assessment workflow applying to our fake academic website.

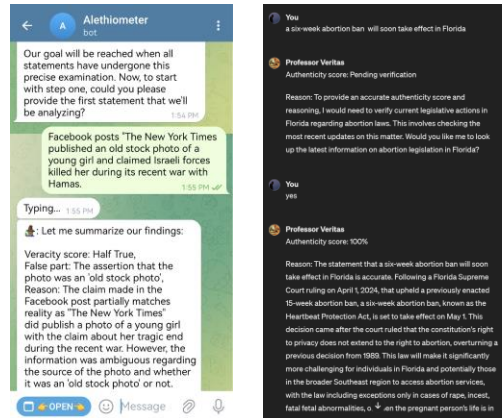
Figure 6: Screenshots of two veracity assessment chatbots: Telebot on Telegram (right) and Chatbot via GPTs (left).

## Testing on Digital Content

Figure 4 illustrates how we manually apply this workflow to a specific example: President Trump's farewell speech[7]. This speech was widely criticized for inaccuracies and exaggerations[8]. In this analysis, Trump's six-sentence statement received a Global Veracity score of 63%, with errors in each sentence underlined in red.

When applied on our own fabricated academic website, as shown in Figure 5, the Global Veracity score was 45%. False statements are underlined in red, and a sentence with a 0% veracity score is highlighted in blue.

Thus, the workflow identifies inaccuracies in our proposed everyday content and assigns an impartial score. We see it as a promising solution to the challenges outlined in our prior practice, designed to enhance the public's capacity to distinguish truth in the era of Generative AI.

## Application on Chatbots

We created two agent chatbots utilizing our workflow, enabling users to perform veracity assessments via conversation. The first, @Alethiometer (Figure 6, left), is readily accessible on Telegram, while the second is available through GPT platforms (Figure 6, right). Users can present statements to these bots, which then analyze, score for veracity, identify inaccuracies, and elucidate the basis of their evaluations. Figure 6 displays conversation screenshots with both bots.

## Conclusions and Future Works

We acknowledge ethical concerns and the current research gap in generative AI study. Through our art practice, we examine human reactions and the implications of generated fake content in real life.

Through this art practice, we have highlighted a critical issue: generative AI significantly eases the creation of misleading information, presenting a growing challenge as the technology evolves.

Then we shift our focus to suggesting future improvements. Our aim was to develop a practical approach to improve public discernment of such content, utilizing the technologies behind its creation. We propose a solution that employs large language models for reasoning, moving beyond traditional binary detection methods.

---

[7] https://www.youtube.com/watch?v=6h5_d3DUdR4

[8] https://reurl.cc/or0MnQ

We have created a workflow to evaluate the veracity of everyday content, identifying inaccuracies through manual testing on videos and web pages. This led to the development of chatbots that assist in recognizing falsehoods, thereby increasing vigilance.

Looking forward, we plan to enhance our workflow's accuracy and envision speculative fact-checking wearables, such as GPT-equipped reading glasses and clip-on devices, to verify spoken information's accuracy. These glasses could also assess visual content's veracity, displaying authenticity scores on their lenses.

Moreover, balancing technological progress with ethical standards and public awareness is essential. Collaborative efforts across fields like computer science, sociology, psychology, and law are vital to tackle these complex challenges. Our ultimate aim is to create an environment where information can be easily verified, empowering the public to trust the content they encounter daily.

## Acknowledgements

The text in this manuscript was grammar-checked by ChatGPT4. The text in Figure 1 and the reasons in Table 2 are generated by ChatGPT4. Figure 6 was initially created using DALL-E and subsequently modified by the author.